\documentclass[lettersize,journal]{IEEEtran}
\usepackage{amsmath,amsfonts}
\usepackage{algorithmic}
\usepackage{algorithm}
\usepackage{array}
\usepackage{textcomp}
\usepackage{stfloats}
\usepackage{url}
\usepackage{verbatim}
\usepackage{graphicx}
\usepackage{cite}
\usepackage{subfigure}
\begin{document}

\title{CCAT: LED Mapping and Characterization of the 280 GHz TiN KID Array}

\author{Alicia Middleton, Steve K. Choi, Samantha Walker, Jason Austermann, James R. Burgoyne, Victoria Butler, Scott C. Chapman, Abigail T. Crites, Cody J. Duell, Rodrigo G. Freundt, Anthony I. Huber, Zachary B. Huber, Johannes Hubmayr, Ben Keller, Lawrence T. Lin, Michael D. Niemack, Darshan Patel, Adrian K. Sinclair, Ema Smith, Anna Vaskuri, Eve M. Vavagiakis, Michael Vissers, Yuhan Wang, Jordan Wheeler

\thanks{A. Middleton, V. Butler, A. T. Crites, C. J. Duell, Z. B. Huber, B. Keller, L. T. Lin, M. D. Niemack, D. Patel, E. Smith, and Y. Wang are with the Department of Physics, Cornell University, Ithaca, NY 14853, USA. (e-mail: amm542@cornell.edu).

S. K. Choi is with the Department of Physics and Astronomy, University of California, Riverside, CA 92521, USA.

S. Walker is with the Department of Physics, Cornell University, Ithaca, NY, 14853 USA and the Cornell Center for Materials Research, Ithaca, NY, 14853 USA.

J. Austermann, J. Hubmayr, M. Vissers, and J. Wheeler are with the Quantum Sensor Division, National Institute of Standards and Technology, Boulder, CO 80305, USA.

A. Vaskuri is with the Quantum Sensor Division, National Institute of Standards and Technology, Boulder, CO 80305, USA and the University of Colorado, Boulder, CO 80305, USA.

J. R. Burgoyne and A. K. Sinclair are with Department of Physics and Astronomy, University of British Columbia, Vancouver, BC V6T 1Z4, Canada.

S. C. Chapman is with the Department of Physics and Astronomy, University of British Columbia, Vancouver, BC, V6T 1Z4, Canada and the Department of Physics and Atmospheric Science, Dalhousie University, Halifax, NS B3H 4R2, Canada.

R. G. Freundt is with the Department of Astronomy, Cornell University, Ithaca, NY 14853, USA.

A. I. Huber is with the Department of Physics and Astronomy, University of Victoria, Victoria, BC V8P 5C2, Canada.

E. M. Vavagiakis is with the Department of Physics, Duke University, Durham, NC 27710, USA and the Department of Physics, Cornell University, Ithaca, NY 14853, USA.
}}



\maketitle

\begin{abstract}
Prime-Cam, one of the primary instruments for the Fred Young Submillimeter Telescope (FYST) developed by the CCAT Collaboration, will house up to seven instrument modules, with the first operating at 280 GHz. Each module will include three arrays of superconducting microwave kinetic inductance detectors (KIDs). The first KID array fabricated for the 280 GHz module uses titanium-nitride (TiN) as the superconducting material and has 3,456 individual detectors, while the other two arrays use aluminum. This paper presents the design and laboratory characterization of the 280 GHz TiN array, which is cooled below its critical temperature to $\sim$0.1 K and read out over six RF feedlines.  LED mapping, a technique for matching the measured resonant frequency of a detector to its physical position, was performed on the array so that the results can be used to lithographically trim the KID capacitors and increase the yield of the array by reducing frequency collisions.  We present the methods and results of LED mapping the 280 GHz TiN KID array before deployment on FYST.
\end{abstract}


\section{Introduction}

\begin{figure}[t!]
    \centering
    \begin{subfigure}
        \centering
        \includegraphics[width=3.5in]{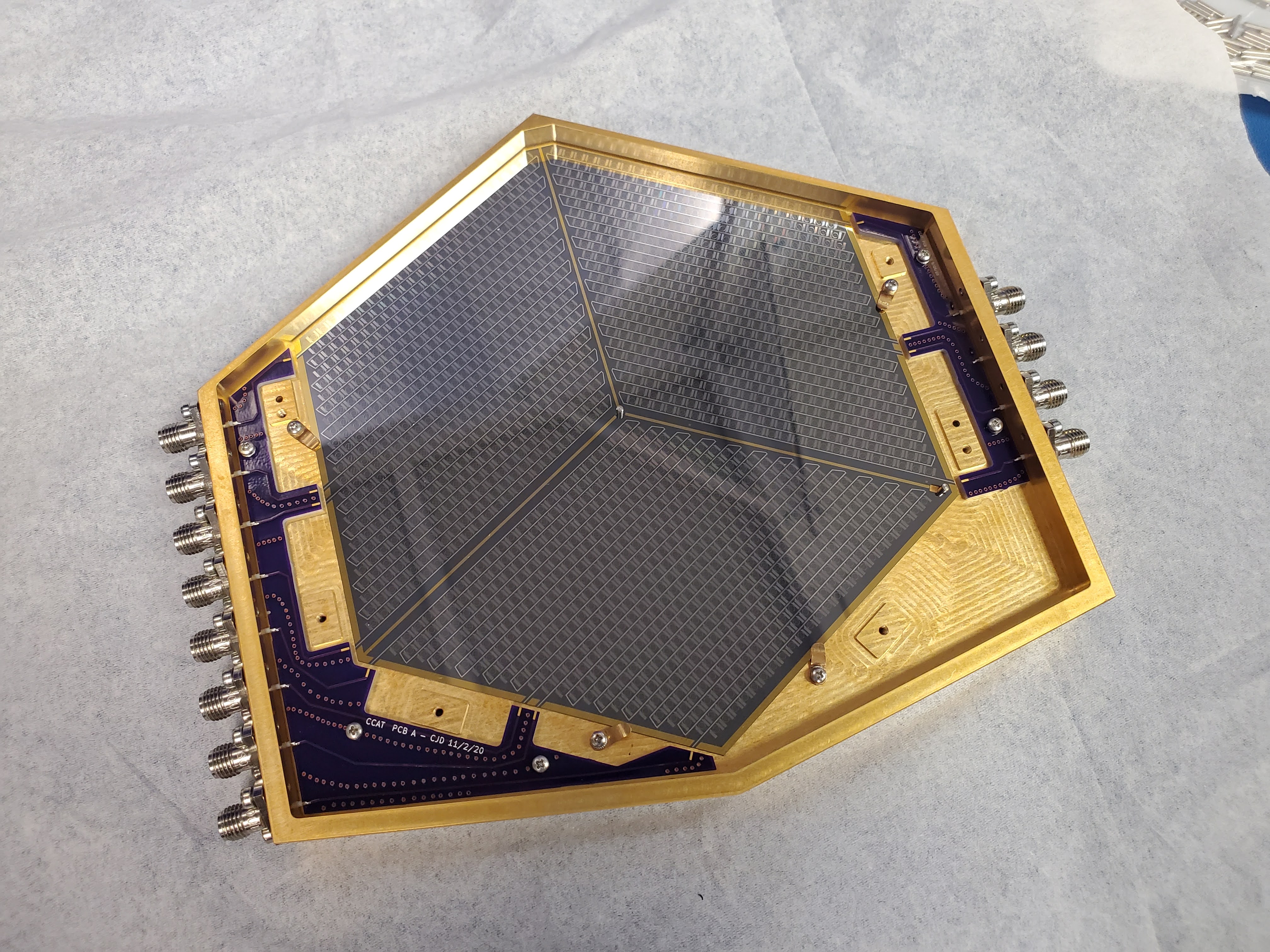}
    \end{subfigure}%

    \begin{subfigure}
        \centering
        \includegraphics[width=3.5in]{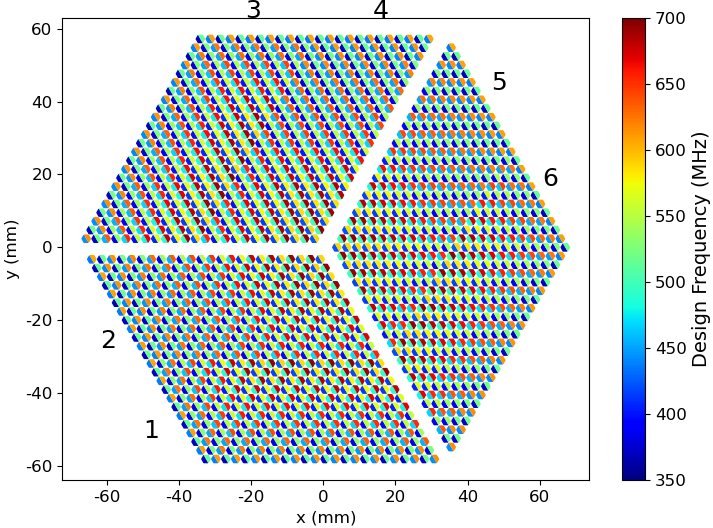}
    \end{subfigure}
    \caption{Top: The 280 GHz TiN array assembled onto a gold-plated aluminum base.  The array was designed with 3,456 individual detectors and is read out over six RF feedlines.  This figure is adapted from \cite{TiNcharacterization}. Bottom: Designed frequencies for each detector.  Each pixel has two individual detectors for polarization sensitivity, with their resonant frequencies designed in an alternating pattern between lower and higher frequencies to minimize crosstalk. The sections of the array are labeled by their feedline numbered 1-6. }
    \label{fig:array}
\end{figure}

The Fred Young Submillimeter Telescope (FYST) is a six-meter-aperture submillimeter telescope to be located at a 5,600~m elevation site in Cerro Chajnantor.  This outstanding high-altitude site will allow for high atmospheric transmission in the submillimeter to millimeter wavelength range that will be measured using the Prime-Cam instrument, which is one of the primary receivers for FYST and is currently being constructed for use by the CCAT Collaboration \cite{CCAT}.

Prime-Cam will house up to seven instrument modules, including those for broadband observations ranging from 280$\textendash$850 GHz as well as imaging spectrometers for line intensity mapping \cite{primecam1} \cite{primecam2}.  Prime-Cam's multi-frequency capabilities will enable studies of science goals including Big Bang cosmology through the Epoch of Reionization, measuring cosmic microwave background (CMB) foregrounds, and tracing galaxy cluster evolution with the Sunyaev-Zeldovich effect \cite{CCAT}.

For first light, the 280 GHz module will be deployed with Mod-Cam \cite{modcam}, and will later be moved to the Prime-Cam instrument for measurements with multiple frequency modules.  The focal plane of the 280 GHz instrument module will consist of three kinetic inductance detector (KID) arrays with a total of 10,352 detectors. One of the arrays uses a TiN-Ti-TiN trilayer as the superconducting material, while two use Al.  The TiN array was the first array fabricated for CCAT, with Al selected later for the other two arrays to improve $1/f$ noise performance \cite{modcam}.  

This paper focuses on the 280 GHz TiN array.  In the following section, we describe the design of the 280 GHz TiN KID array and the experimental setup, and then the process used for LED mapping the array.  We then discuss the results from this LED mapping process and describe the next steps for the TiN array.

\section{Instrument and Experiment}

A full description of the design and assembly of the 280 GHz TiN array is provided in \cite{280GHzdesign}.  The TiN array uses a TiN/Ti/TiN trilayer as its superconducting material, which has a critical temperature of $T_c \sim$1.1 K \cite{TiNcharacterization}. 

The array was designed with a total of 3,456 individual detectors that are read out over six RF feedlines with 576 detectors per feedline.  There are two detectors per pixel which are rotated such that they are orthogonal to one another for polarization sensitivity.  The designed resonant frequencies for each detector are shown in the bottom panel of Figure \ref{fig:array}.  

The array was assembled onto a gold-plated aluminum base, pictured in the top panel of Figure \ref{fig:array}, and coupled to an aluminum feedhorn array.  This array package was then mounted to a Bluefors LD400 system and cooled to a base temperature of $\sim$100 mK,  and read out using a second generation Reconfigurable Open Architecture Computing Hardware (ROACH-2) board \cite{ROACH2}.  

\begin{figure}[t!]
    \centering
    \begin{subfigure}
        \centering
        \includegraphics[width=3in]{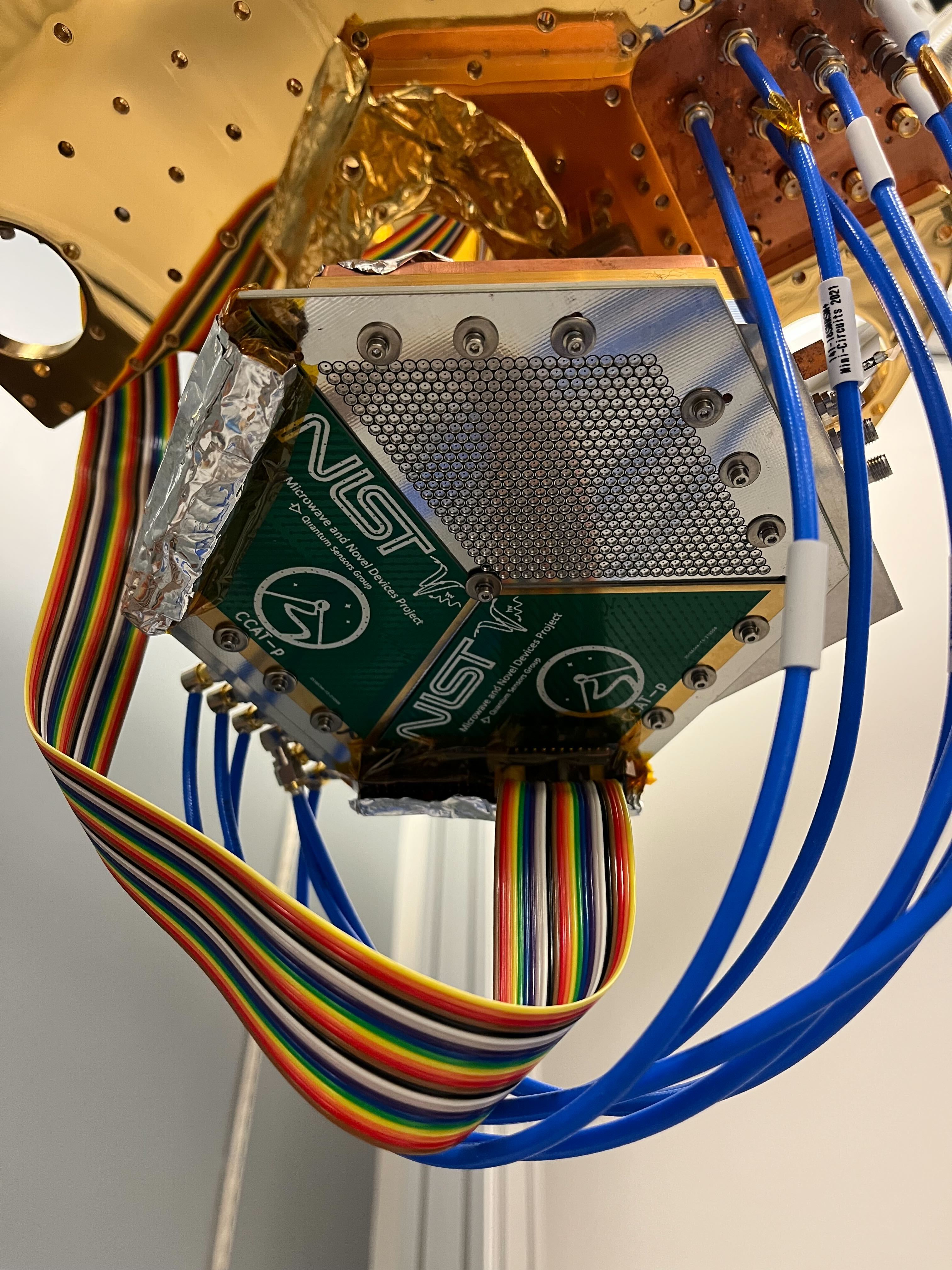}
    \end{subfigure}%

    \begin{subfigure}
        \centering
        \includegraphics[width=3in]{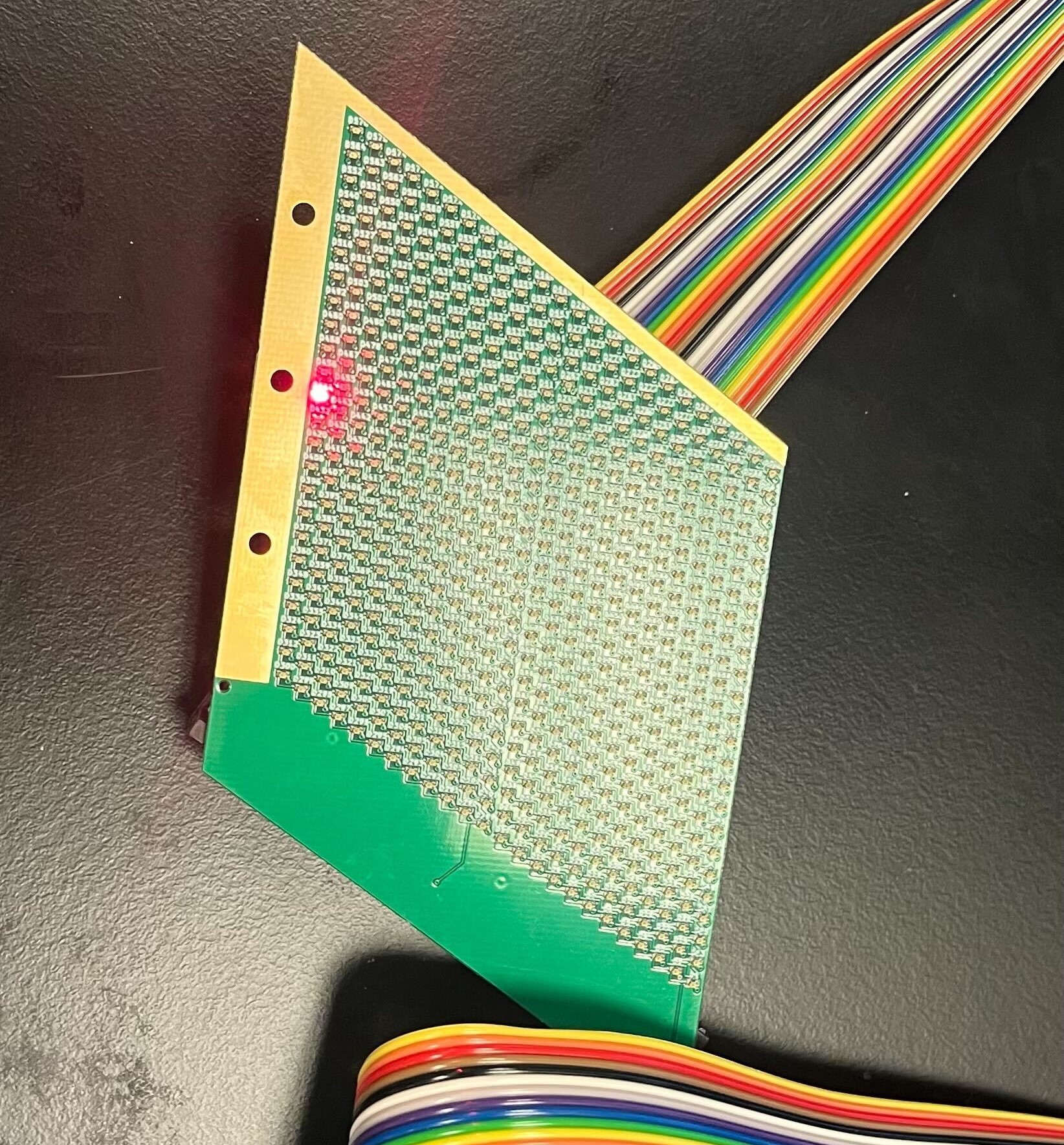}
    \end{subfigure}
    \caption{Top: The LED mapper PCB interfaced to the feedhorns coupled to the array, which was assembled onto the gold-plated aluminum base shown in Figure \ref{fig:array} and mounted to a Bluefors LD400. A second PCB that is not powered was installed to ensure the interface stays level to the feedhorn array.  Bottom: The side of the LED mapper PCB which faces the array, with one LED turned on.  The LED mapper was designed to line up with each third of the array, with an LED positioned over each feedhorn.}
    \label{fig:LEDmapper}
\end{figure}

\begin{figure*}
\label{fig:timestream}
\includegraphics[width = \textwidth]{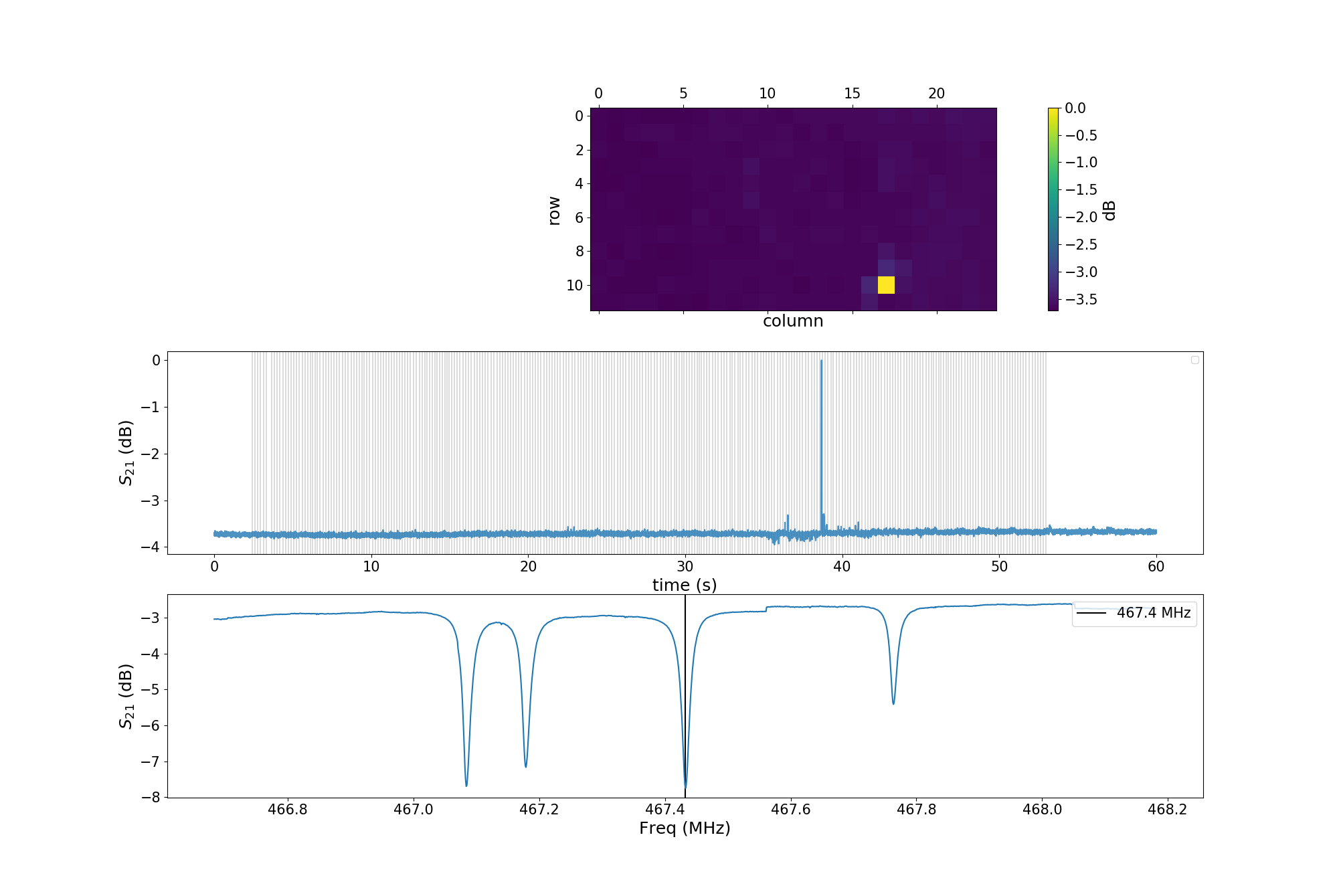}
\caption{Visualization of the LED mapping process. First, we measured an $S_{21}$ sweep of a feedline to find the resonant frequencies for all detectors on the feedline.  A section of this $S_{21}$ trace is shown in the bottom panel.  The resonant frequency for this particular detector is marked with a solid black line, at $f_0 = 467.4$ MHz.  The middle panel shows the timestream for this resonator taken while sequentially activating the LEDs across the LED mapper.  The signal from an LED causes $f_0$ and $Q$ of a resonator to shift, which leads to a sharp peak in the timestream.  The thin gray lines in the middle plot correspond to the timestamps of the activation of each LED.  The top panel is a visualization of the half of the LED mapper covering the detectors on the feedline we were measuring, with the colorbar giving the maximum signal measured after the timestamp corresponding to the row and column of the LED that was activated at the time.  For example, the yellow square at (10, 17) in the top panel, which indicates a high amount of signal compared to the signal measured after other LEDs, corresponds to the peak in the timestream in the middle panel.  Using this information, we then used the orientation of the LED mapper with respect to the array to determine the physical location of the detector corresponding to the resonator found in the $S_{21}$ trace.}
\end{figure*}

Although the resonant frequencies $f_0$ of the detectors across an array can be identified by locating the resonators in an $S_{21}$ sweep, it is less trivial to identify where these detectors are physically located on the array itself.  LED mapping is a technique used to match the measured $f_0$ of each detector to its physical location, which is essential information for maximizing the yield of a KID array.  

The resulting LED resonator-position map from LED mapping is used to lithographically trim the capacitors of each detector to reduce frequency collisions, leading to an improved array yield. Frequency collisions, which occur when the peaks of two or more resonators overlap in frequency space, reduce sensitivity of the array by preventing differentiation of individual detectors.  This capacitor trimming process, which will be performed at the National Institute of Standards and Technology (NIST) in Boulder, CO, is described in \cite{capacitortrimming}.

When a KID absorbs incident photons from an LED, Cooper pairs are broken in the superconducting material, generating quasiparticles.  This causes an increase in the kinetic inductance of the resonator and results in a downward shift of the resonant frequency.  Quasiparticle generation also causes an increase in losses that lower quality factor $Q$ of the resonator.  A decrease in $f_0$ and $Q$ are therefore indicative of a KID detecting a signal from an LED.

An LED mapper PCB, pictured in Figure \ref{fig:LEDmapper}, was fabricated to map the measured detector resonant frequencies to their physical locations on the 280 GHz TiN array.  The LED mapper, which has a total of 576 surface-mounted device (SMD) LEDs of type 0402 (1005 metric), was designed to line up with each third of the array, with an LED positioned over the feedhorn for each pixel. It was coupled to the feedhorn array via an interface as shown in the top panel of Figure \ref{fig:LEDmapper}.  The LED mapper is addressed using an Adafruit FT232H breakout board which enables the activation of an LED in any particular location for any length of time, and allows us to blink through the LEDs in any order while saving the timestamps of when the LED at each position is turned on. 

Using the ROACH-2, we first conducted an $S_{21}$ sweep of a feedline to identify the resonant frequencies of each detector.  We then recorded timestreams of the $S_{21}$ values of these resonators while sequentially activating the LEDs across the LED mapper, which results in sharp peaks in the timestreams when the LED corresponding to the physical location of a resonator is activated due to the aforementioned decrease in $f_0$ and $Q$.  Lastly, we combined the timestream data of each detector with the timestamps of the LEDs to match the timestream peaks with the corresponding LED location.  A visualization of this LED mapping process for one detector is presented in Figure 3.

\section{Results}

\begin{figure}[h!] 
    \includegraphics[width= 3.5in]{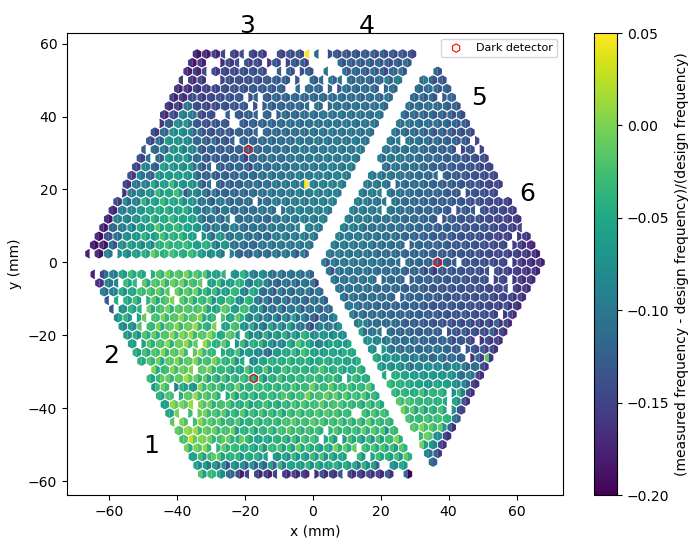}
    \caption{The results of LED mapping the 280 GHz TiN array.  The yield for each network is given in Table \ref{tab:yield}.  The color scale gives the difference between measured and designed resonant frequencies for each detector, of which there is a spread of $\sim$20$\%$ across the array.  White detector locations represent physical positions for which a resonator was not found in the $S_{21}$ sweep of the feedline and therefore could not be mapped.  Mapped resonators which are collided in frequency space are included in this plot. The sections of the array are labeled by their feedline numbered 1-6. }
    \label{fig:results}
\end{figure}

The results from the LED mapping process for the 280 GHz TiN array are shown in Figure \ref{fig:results}.  The plot shows the difference between the designed and measured resonant frequencies for each detector.  The yield of the detectors on each feedline, which is the number of detectors found and mapped compared to the number of detectors designed, is given in Table \ref{tab:yield}.

\begin{table}[!t]
\caption{Total Yield of Each Feedline\label{tab:table1}}
\label{tab:yield}
\centering
\def\arraystretch{1.5}%
\begin{tabular}{|c|c|c|}
\hline
Feedline & Detectors Mapped & Yield\\
\hline
1 & 544 & $\sim$95$\% $ \\
\hline

2 & 548 & $\sim$95$\% $\\
\hline
3 & 537 & $\sim$93$\% $\\
\hline
4 & 543 & $\sim$94$\% $\\
\hline
5 & 557 & $\sim$96$\% $\\
\hline
6 & 553 & $\sim$96$\% $\\
\hline

\end{tabular}
\end{table}

Between $\sim$93$\%$ and $\sim$96$\%$ of the designed detectors were found by the $S_{21}$ sweep and mapped for each feedline.  As can be seen by the color scale in Figure \ref{fig:results}, there is a spread of $\sim$20$\%$ in the variation of the measured frequencies from the designed frequencies.  The detector locations that are white in the plot represent positions for which a resonator was not mapped.  These detectors were not found by the $S_{21}$ sweep and are missing due to fabrication defects.  Resonators that are collided in frequency space are included in the total yield of the array.

The post-LED mapping lithographic capacitor trimming process has already been completed at NIST on one of the Al KID arrays for the 280 GHz instrument module.  Before trimming, the total yield of the Al array was found to be $\sim$97$\%$ and a spread of $\sim$20$\%$ from the designed resonant frequencies was observed, similar to that of this TiN array.  The number of collided resonators, which is defined as the number of resonators within five linewidths of another resonator, was reduced by more than 30$\%$ after trimming. Laboratory measurements of the Al array performance, including LED mapping and capacitor trimming results, will be presented in an upcoming paper.  Capacitor trimming of the TiN array based on the LED mapping results presented in this paper will be completed at NIST soon.

\section{Conclusion}

Using custom LED boards, we successfully mapped 3,282 resonator positions, which is $\sim$95$\%$ of the CCAT 280 GHz TiN KID array. This is one of the largest KID arrays to undergo LED mapping to date. This LED mapping work informs the post-editing process designed to reduce resonator collisions and represents a significant step in maximizing KID array yields.  

Testing will continue for the 280 GHz instrument module in preparation for the deployment of the 280 GHz KID arrays in Mod-Cam with a first light target of early 2026.  Deployment of the 280 GHz module in Prime-Cam, along with instrument modules operating at other frequencies, is planned for the following year.

\section{Acknowledgements}

The CCAT project, FYST and Prime-Cam instrument have been supported by generous contributions from the Fred M. Young, Jr. Charitable Trust, Cornell University, and the Canada Foundation for Innovation and the Provinces of Ontario, Alberta, and British Columbia. The construction of the FYST telescope was supported by the Gro{\ss}ger{\"a}te-Programm of the German Science Foundation (Deutsche Forschungsgemeinschaft, DFG) under grant INST 216/733-1 FUGG, as well as funding from Universit{\"a}t zu K{\"o}ln, Universit{\"a}t Bonn and the Max Planck Institut f{\"u}r Astrophysik, Garching.

\vfill

\end{document}